\documentclass[sn-mathphys,Numbered]{sn-jnl}

\usepackage{graphicx}%
\usepackage{multirow}%
\usepackage{amsmath,amssymb,amsfonts}%
\usepackage{amsthm}%
\usepackage{mathrsfs}%
\usepackage[title]{appendix}%
\usepackage{xcolor}%
\usepackage{textcomp}%
\usepackage{manyfoot}%
\usepackage{booktabs}%
\usepackage{algorithm}%
\usepackage{algorithmicx}%
\usepackage{algpseudocode}%
\usepackage{listings}%
\usepackage{orcidlink}
\usepackage{gensymb}

\usepackage{MnSymbol}
\usepackage{color,soul} 

\newcommand{\mat}[3]{\displaystyle\left\langle #1 \left| #2 \right| #3 \right\rangle}

\newcommand{\ket}[1]{\displaystyle\left| #1 \right\rangle}

\newcommand{\braket}[2]{\displaystyle\left\langle #1|#2\right\rangle}

\newcommand\apj{{Astrophys Journal}}
\newcommand\apjl{{Astrophys Journal Letters}}
\newcommand\prl{Phys. Rev. Letters}
\newcommand\pra{Phys Rev A}
\newcommand\pre{Phys Rev E}
\newcommand\aap{Astron \&Astrophysics}
\newcommand\mnras{MNRAS}
\newcommand\araa{Ann Rev of Astron and Astrophys}
\newcommand\nat{Nature}

\theoremstyle{thmstyleone}%
\theoremstyle{thmstyletwo}%

\theoremstyle{thmstylethree}%

\begin{document}

\title{Enhancements of Electron-Atom Collisions due to Pauli Repulsion in Neutron-Star Magnetic Fields}

\author*[1,2,5]{\sur{Gomez}, \fnm{T.~A.}  \orcidlink{0000-0001-8748-5466}}\email{thomas.gomez@austin.utexas.edu}

\author[3]{\sur{Zammit}, \fnm{M.~C.}  \orcidlink{0000-0003-0473-379X}}
\equalcont{These authors contributed equally to this work.}

\author[4]{\sur{Bray}, \fnm{I.} \orcidlink{0000-0001-7554-8044}}
\equalcont{These authors contributed equally to this work.}

\author[3]{\sur{Fontes}, \fnm{C.~J.} \orcidlink{0000-0003-1087-2964}}
\equalcont{These authors contributed equally to this work.}

\author[1]{\sur{White}, \fnm{J.}  \orcidlink{0000-0003-4052-2746}}
\equalcont{These authors contributed equally to this work.}

\affil[1]{\orgdiv{Department of Astronomy}, \orgname{University of Texas at Austin}, \city{Austin}, \state{Texas}, \postcode{78712}, \country{USA}}
\affil[2]{\orgdiv{Department of Astrophysical and Planetary Sciences}, \orgname{University of Colorado}, \city{Boulder}, \state{Colorado}, \postcode{80305}, \country{USA}}

\affil[3]{\orgname{Los Alamos National Laboratory}, \city{Los Alamos}, \state{New Mexico}, \postcode{87545}, \country{USA}}

\affil[4]{\orgdiv{Department of Physics and Astronomy}, \orgname{Curtin University}, \city{Perth}, \state{Western Australia}, \postcode{6845}, \country{Australia}}

\affil[5]{Hale Fellow}

\abstract{
Neutron star surfaces and atmospheres are unique environments that sustain the largest-known magnetic fields in the universe \cite{NSbook}. 
Our knowledge of neutron star material properties, including the composition and equation of state, remains highly unconstrained. 
Electron-atom collisions are integral to theoretical thermal conduction and spectral emission models that describe neutron star surfaces \citep{Potekhin99,Bergeron92}. 
The theory of scattering in magnetic fields was developed in the 1970s, but focused only on bare nuclei scattering \citep{Ventura73}.
In this work, we present a quantum treatment of atom-electron collisions in magnetic fields; of significant importance is the inclusion of Pauli repulsion arising from two interacting electrons \citep{Oppenheimer28,Bray92}.
We find strange behaviors not seen in collisions without a magnetic field.
In high magnetic fields, Pauli repulsion can lead to orders of magnitude enhancements of collision cross sections.
Additionally, the elastic collision cross sections that involve the ground state become comparable to those involving excited states, and states with large orbits have the largest contribution to the collisions.
We anticipate significant changes to transport properties and spectral line broadening in neutron star surfaces and atmospheres, which will aid in spectral diagnostics of these extreme environments.
}





\maketitle


Neutron stars are truly unique objects, characterized by extreme states of matter seen nowhere else in the universe.
They can reach densities up to $10^{14} ~\mathrm{g/cm}^3$ in their interiors and can sustain magnetic fields from $10^9$~G to $10^{15}$~G in magnetars \cite{NSbook}.
It is largely unknown what the state of matter is at such extreme density. 
Some predictions include neutron matter or even quark matter, each having their own equation of state~\citep{Lattimer01}.
One proposed method for constraining the equation of state (in the absence of experiments) is by accurately determining surface conditions, i.e. the mass and radius, of neutron stars~\citep{Lindblom92}.
Much effort has therefore been invested in trying to determine the masses and radii of neutron stars~\citep{Heinke14,Ozel16,NSbook,NICER1,NICER2}, though the results are still accompanied by significant uncertainties.

The surface conditions of neutron stars are less extreme than the interiors, but are still exotic compared to other stellar surfaces because they sustain magnetic fields above $10^{9}$~G---the only known objects in the universe known to do so.
In these extreme magnetic fields, atoms no longer have spherical symmetry and instead tend toward cylindrical symmetry \citep{Ruder94}.
To better understand the surface conditions of neutron stars, we need to understand fundamental atomic processes, in particular atomic collisions.
Collisions directly affect transport properties such as thermal and electrical conductivity~\citep{Ziman54}.
Transport is significantly affected in directions perpendicular to the magnetic field, resulting in 180\degree ~scattering \citep{Baalrud17}, though transport along the magnetic field is expected to be unaffected compared to their field free values \citep{Braginskii,LeeMore84}.
Further, collisions inform spectral line shapes~\citep{Baranger58c,Fano63,Gomez22}, which could be one of the tools to determine surface parameters from spectra \citep{Gomez23} in the same manner as is currently done with white dwarfs \citep{Bergeron92}.
Such determinations will be more feasible with future planned X-ray missions, such as XRISM \citep{XRISM} and ARCUS~\citep{Smith17_arcus}.

The theory of atomic collisions in magnetic fields is severely under-developed compared to collisions without a magnetic field.
It is important to note that current calculations of particle dynamics for transport \citep{Baalrud17} and line shapes \citep{Ferri21} treat the plasma particles classically.
If quantum behavior of the plasma particles is included, then only scattering off of bare nuclei is considered \citep{Ventura73,Lafleur19}.
\citet{Gomez23} performed the first quantum-mechanical treatment of electron-atom scattering in the context of spectral line broadening, but neglected to include the important Pauli repulsion between colliding electrons.
Pauli repulsion accounts for the fact that electrons are indistinguishable fermions, which is often referred to as the Pauli exclusion principle.
The impact of Pauli repulsion on electron-atom collisions in magnetic fields is unknown, but is expected to be different due to the change in the geometry of the scattering problem.

Since the early work of~\citet{Oppenheimer28}, Pauli repulsion has been known to be an important physical effect that needs to be included in scattering calculations.
Due to high fidelity experiments, Pauli repulsion plays an essential role in reproducing measurements, though only making (relatively) modest changes to the scattering transition $T$-matrix elements, never resulting in changes of more than factors of two.
However, as we will show, in strong magnetic fields, Pauli repulsion becomes the dominant contribution to electron-atom collisions.

\begin{figure}[h!]
\centering
\includegraphics[scale=0.45,trim=0cm 0 0 0]{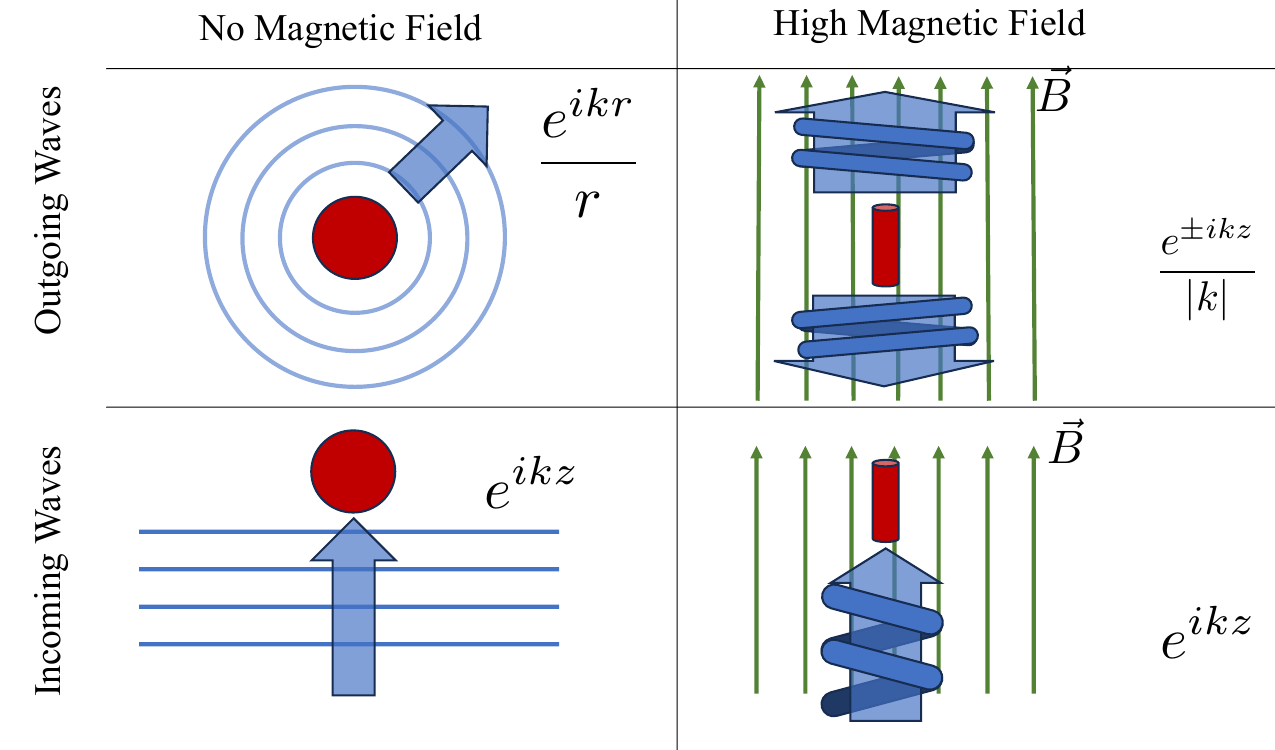}
\caption{Illustration of scattering processes with (right) and without (left) a magnetic field (field lines indicated by green arrows). The target atom is shown in red with the magnetic field making the target cylindrical. The scattering process is separated into the incoming (bottom) and outgoing (top) waves, which can be thought of as the waves before and after the collision, respectively. Without a magnetic field, scattering electrons have the full three-dimensional degrees of freedom. In the presence of a magnetic field, the Lorentz force confines the motion of electrons perpendicular to the magnetic field, but electrons can still freely propagate and scatter parallel to the field. Further, after a collision, electrons can also change their orbits in the magnetic field.}
\label{cartoon}
\end{figure}

Electron-atom collisions in a large magnetic field more closely resembles concentric cylinders rather than spheres, as illustrated in figure~\ref{cartoon}.
Without a magnetic field, the picture of scattering system consists of an incoming plane wave and an outgoing spherical wave.
In a magnetic field, the magnetic field significantly alters the trajectories of the projectile electrons, limiting the free motion perpendicular to the field, while still propagating freely parallel to the field.
This effectively reduces scattering in a magnetic field to a one-dimensional problem, rather than the usual three-dimensional problem.
In this effective 1D problem, we have an incoming wave and outgoing reflected and transmitted waves.

The elemental composition of neutron stars is unknown.
Predictions of the composition of non-accreting neutron star atmospheres range from light elements, such as hydrogen or helium \citep{Ho03,Guver11},
to mid-Z elements, such as carbon, oxygen, or neon~\citep{Mori06,Ho09,Alford23}.
Therefore, for this study, we will examine how magnetic fields affect electron collisions of both neutral hydrogen and highly-ionized oxygen.

The Hamiltonian for a free particle in a magnetic field (which is conventionally aligned in the $z$ direction) contains a harmonic oscillator potential in the $x$-$y$ direction \citep{Ventura73,Clark83},
\begin{equation}
    H_0^{{p}} = -\frac{1}{2}\nabla^2 + \frac{1}{2}L_z\beta + \frac{1}{2}\left(\frac{\beta
    }{2}\right)^2\varrho^2 + S_z\beta + V_{\mathrm{nuc}}(z,\varrho),
\end{equation}
where we have used atomic units throughout ($m_e=\hbar=e=a_0=1$ and $\beta=B/B_0$ where $B_0=2.35\times10^9$~G), the superscript ${p}$ denotes the ``projectile" in the scattering problem, and the subscript ``0" denotes that the particle is not interacting with other electrons.
Here, $L_z$ and $S_z$ are orbital and spin angular momentum operators, and $V_{\mathrm{nuc}}(z,\varrho)$ is a long-range nuclear potential.

The only direction where electrons can freely propagate is the $z$ direction.
The magnetic field states and energies are described by 
\begin{align}
    \braket{\vec{r}}{knm} &= N e^{ikz}e^{-\beta \varrho^2/4}\left(\frac{\beta}{2}\varrho\right)^{|m|/2} L^{|m|}_{n}(\beta\varrho^2/2) e^{im\varphi} \label{waves}\\
    &~~\mathrm{and}~~\nonumber\\
    \varepsilon_{knm} &= \frac{1}{2}k^2 + \frac{\beta}{2}(2n+|m|+m+1) + m_s\beta,
\end{align}
respectively, where $k$ is the momentum in the $z$ direction, $n$ is the ``Landau" number, and $m$ is the azimuthal/magnetic quantum number, defined the same way as in the field-free case.
Here, $N$ is a normalizing factor, and $L^{m}_n(x)$ is an associated Laguerre polynomial.
The presence of the magnetic field creates wavefunctions that decay to zero as $\varrho$ becomes large, i.e. the wavefunctions are bound.
This means that electrons can freely scatter is the $z$ direction as well as change to a different cylindrical orbit, i.e. scatter into states of different $n$ or $m$ quantum numbers.

Fortunately, the formulation for scattered waves is unchanged due to the presence of a magnetic field. The difference is that the set of wavefunctions to construct the scattered wave are those in Eq \eqref{waves} instead of three-dimensional plane waves.
The scattered wave is defined as~\citep[][]{Bray92}
\begin{equation}
    \ket{f_{j,i}} = \delta_{j,i}\ket{k_in_im_i} + \sumint_jdk_j \frac{\ket{k_jn_jm_j}\mat{k_jn_jm_j \phi_j}{T(E)}{\phi_i k_in_im_i}}{E - \varepsilon_j-\varepsilon_{k_jn_jm_j}},
\end{equation}
where $\phi_i$ is a set of atomic states that the projectile electron is scattering from, $\varepsilon$ denotes energy, and $T(E)$ is the collision ``transition" or $T$-matrix, defined in \citep{LippmannSchwinger50}, which is closely related to the collision amplitude.
We improve over past collision calculations \citep{Gomez23} by including exchange interactions \citep{BetheSalpeter}.
Including exchange results in the important anti-symmetric property that
\begin{equation}
    \braket{\phi_k}{f_{j,i}} = (-1)^S\braket{\phi_j}{f_{k,i}},
\label{antisymmetric}
\end{equation}
where $S$ is the total spin of the atom+projectile system; $S=0$ corresponds to singlet scattering where the spins of the particles are anti-symmetric, and $S=1$ is triplet scattering where the spatial wavefunctions are anti-symmetric.
To properly impose the boundary condition in Eq \eqref{antisymmetric}, we follow \citet{Bray92} by modifying the 
Pauli repulsion terms in the atom-projectile interaction\footnote{See \citet{Gomez23} for direct terms.}
\begin{equation}
   (-1)^S[H-E]P \Rightarrow
   (-1)^S[H-(1-\theta)E]P - \theta EI_p,
\end{equation}
where $H$ and $E$ are the total Hamiltonian and energy, respectively, and $P$ is the  ``Pauli" exchange operator.  For the new form, $I_p$ is a unit operator acting on the projectile electron space, and $\theta$ is an arbitrary non-zero scalar where the $T$-matrix solutions are independent of the value of $\theta$.\footnote{We note that in low-magnetic field cases, $|\theta|$ needs to be large to satisfy Eq \eqref{antisymmetric}. We think this is required due to the competing symmetries: at low fields, the target atom is more spherical, but the projectile is still cylindrical. At high magnetic fields seen in neutron stars, such as $10^{12}$~G, both target and projectile are cylindrical and nearly any non-zero value of $\theta$ can be used.}

\begin{figure}[h!]
\includegraphics[scale=0.75,trim=0cm 0 0 0]{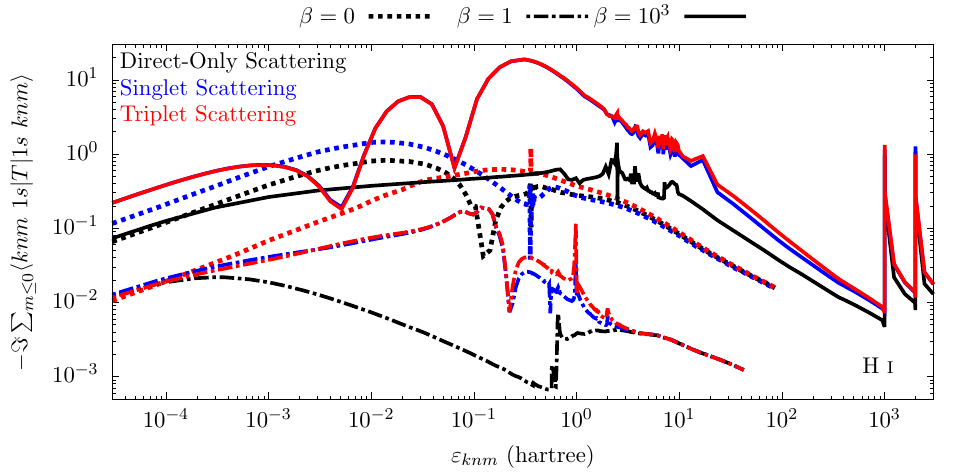}
\caption{Illustration of the importance of exchange on the collision $T$-matrices of the $1s$ state of neutral hydrogen in high magnetic fields. We have three cases: zero field 
(dotted line), $\beta$=1 (dot-dashed line), and $\beta=10^3$ (solid line). For these three cases, we show the imaginary part of the elastic $T$-matrix as a representation of the total cross section through the optical theorem.
We perform three types of calculations: direct-only, singlet scattering and triplet scattering. At $\beta=1$, the $T$-matrices decrease, but as the field goes up, then the $T$-matrices increase over their field-free values.}
\label{Hcollisions}
\end{figure}

In figure \ref{Hcollisions}, we compare elastic $T$-matrices of the $1s$ state of neutral hydrogen for $\beta=0,~1,~\mathrm{and}~10^3$ with different physics included.
These conditions were chosen to sample a range of relative importance between Coulomb and magnetic interactions.
Clearly, at $\beta=0$, the Coulomb interaction between particles is the only thing that dictates electron motion.
At $\beta=1$, the magnetic energy is of the same order as the nuclear energy; this is at the lower end of the range of magnetic fields expected to be found in neutron stars.
Finally, at $\beta=10^3$, the magnetic energy dominates over the Coulomb interactions and is a ``typical" field found on neutron star surfaces.
Examining neutral hydrogen collisions often serves as a benchmark for atomic collision calculations \citep{MottMassey,Moiseiwitsch68}.
We examine, specifically, the imaginary part of the elastic $T$-matrix, as it is representative of the total cross section through the optical theorem \citep{LippmannSchwinger50,MottMassey,Gomez21}.
We also examine the changes in singlet and triplet scattering (blue and red lines, respectively, in figure \ref{Hcollisions}) from these interactions.
For low and high energies of the projectile, exchange is a modest correction to the $T$-matrices.
However, for $k$ values that correspond to wavelengths on the order of the atom's dimension, Pauli repulsion becomes large compared to direct interactions, and $T$-matrices that include exchange \textbf{\textit{are larger by more than an order of magnitude}} than $T$-matrices that include direct-only interactions.
This increase appears as a broad contribution and is not due to dielectronic resonances.

These significant increases in collisions amplitudes/$T$-matrices are largely due to the confining nature of magnetic fields.
In the absence of a magnetic field, when an electron starts to experience Pauli repulsion, the electron can scatter into a non-zero angle.
But in the presence of a large magnetic field, scattering into an angle away from the $z$ axis is extremely difficult. 
In order to scatter into an excited Landau state, the electron must have a kinetic energy in excess of the magnetic energy.
Therefore, revisiting figure~\ref{cartoon}, when constrained to a particular trajectory, the electron is forced to interact with the atom and its only scattering options are to transmit to the other side, or completely change its direction of propagation in the form of a reflected wave.
Therefore, the electrons are forced to experience the Pauli repulsion, which then becomes the dominant collision process.

\begin{figure}[h!]
\includegraphics[scale=0.75,trim=0cm 0 0 0]{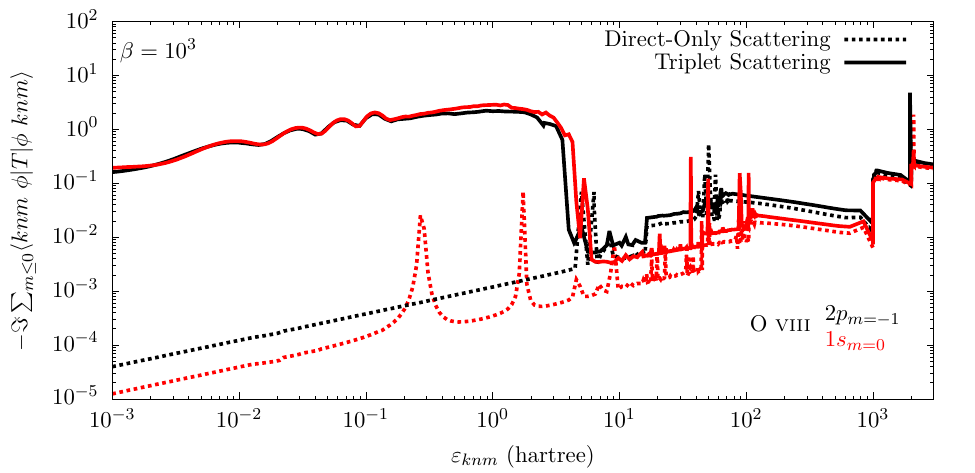}
\caption{Illustration of the impact of Pauli repulsion on the collision $T$-matrices on the $1s_{m=0}$ (red lines) and $2p_{m=-1}$ (black lines) state of H-like oxygen at $\beta=10^3$. The dotted lines neglect Pauli repulsion and the solid lines include it. The most extreme increase in the $T$-matrices is roughly a factor of $10^4$. Pauli repulsion causes the elastic $T$-matrices of the ground and excited states to become comparable.}
\label{Ocollisions}
\end{figure}

There is a stark  difference in the amount of increase seen in neutral hydrogen vs ionized H-like oxygen, the latter is demonstrated in figure \ref{Ocollisions}, where Pauli repulsion leads to increases in the $T$-matrices of \textbf{\textit{4 orders of magnitude}}.
Elastic scattering of ions has an additional complication over neutrals, namely having to include bound states of the projectile \citep{Bray94,Gomez20}.
In H~{\sc i} and even more so in O~{\sc viii}, much of the $T$-matrix enhancement is due to the slow convergence of states with high $m$ values\footnote{States of $m$ up to $150$ were used in figures \ref{Hcollisions} and \ref{Ocollisions}.}; this will be the subject of future study.
Because of the cylindrical symmetry, when $m$ is large, exchange interactions of the type
\begin{equation}
    \ket{\phi_{\nu n_1 m_1} kn_2m_2} \Rightarrow \ket{\phi_{\nu n_2 m_2} kn_1m_1}
\end{equation}
become the dominant type of collision interaction and more $m$ states (both atomic and projectile) are needed to achieve convergence.
The matrix elements for the high partial waves are strongly dependent on the energy of the target atom. 
The energy of the target states can significantly change depending on the mass of the nucleus \citep{Johnson83,PM93}.
And because the $T$-matrix contains energy terms in the exchange part of the interactions, the resulting $T$-matrix solutions are therefore impacted by the mass of the nucleus.

There is further unusual behavior in the collisions of different levels.
In the absence of a magnetic field, the $T$-matrices of excited states, such as $2p$, usually dwarf those of the $1s$ ground state.
This behavior is the reason why line-shape calculations can neglect the collisions of the lower state for $K$-shell transitions \citep{Gomez22}.
This same field-free behavior is repeated in the direct calculation when a magnetic field is present.
But when Pauli repulsion is included, we can see in figure \ref{Ocollisions} that the elastic collisions of $1s$ and $2p_{m=-1}$ are similar at low collision energies.

The present calculations demonstrate that magnetic fields can make Pauli exclusion (or repulsion) become the dominant collision process, resulting in large increases in the elastic collision $T$-matrices, which are representative of the total cross section.
And by large increases, we specifically mean orders of magnitude increases.
These enhancements give some unusual behavior in the $T$-matrix solutions not seen in non-magnetic cases such as the cross section of the ground state having a similar total cross sections (if not a larger cross section) than those of excited states. 
These enhancements will have an effect on two important physical properties: the transport coefficients \citep{Potekhin99} and the spectral line broadening \citep{Gomez23}.
Stronger electron collisions have an inverse relationship with viscosity as well as electrical and thermal conductivity \citep{McQuarrie}, i.e. stronger collisions result in decreased transport properties.
Past work on transport properties, e.g. \citet{Potekhin99}, considered only collision frequencies off bare ion nuclei and neglected any effects from bound electrons.
The work presented here finds that should there be any bound electrons, then transport coefficients could be altered substantially compared to estimates that assume bare nuclei.
Lastly, we conclude that electron broadening would be significantly enhanced in neutron star atmospheres.
This increase may make it so that the plasma broadening is the dominant broadening mechanism, though more work is needed before emergent spectral models are mature enough to use.
The work presented here is an essential aspect of studying the spectra of neutron stars ( atmosphere models can now go beyond the LTE assumption) and---with high enough resolution measurements---allow us to determine the gravity of neutron stars and constrain the equation of state of the interior.

\backmatter





\bmhead{Acknowledgments}

T.A.G. and J.W. acknowledge support from the United States Department of Energy under grant DR-SC0010623 and the Wootton Center for Astrophysical Plasma Properties under the United States Department of Energy collaborative agreement DE-NA0003843.
M.C.Z and C.J.F. acknowledge support from the Los Alamos National Laboratory (LANL) ASC PEM Atomic Physics Project. LANL is operated by Triad National Security, LLC, for the National Nuclear Security Administration of the U.S. Department of Energy under Contract No. 89233218NCA000001.
I. B. acknowledges the Australian Research Council and the support of the National Computer Infrastructure and the Pawsey Supercomputer Centre.


\begin{thebibliography}{41}
\ifx \bisbn   \undefined \def \bisbn  #1{ISBN #1}\fi
\ifx \binits  \undefined \def \binits#1{#1}\fi
\ifx \bauthor  \undefined \def \bauthor#1{#1}\fi
\ifx \batitle  \undefined \def \batitle#1{#1}\fi
\ifx \bjtitle  \undefined \def \bjtitle#1{#1}\fi
\ifx \bvolume  \undefined \def \bvolume#1{\textbf{#1}}\fi
\ifx \byear  \undefined \def \byear#1{#1}\fi
\ifx \bissue  \undefined \def \bissue#1{#1}\fi
\ifx \bfpage  \undefined \def \bfpage#1{#1}\fi
\ifx \blpage  \undefined \def \blpage #1{#1}\fi
\ifx \burl  \undefined \def \burl#1{\textsf{#1}}\fi
\ifx \doiurl  \undefined \def \doiurl#1{\url{https://doi.org/#1}}\fi
\ifx \betal  \undefined \def \betal{\textit{et al.}}\fi
\ifx \binstitute  \undefined \def \binstitute#1{#1}\fi
\ifx \binstitutionaled  \undefined \def \binstitutionaled#1{#1}\fi
\ifx \bctitle  \undefined \def \bctitle#1{#1}\fi
\ifx \beditor  \undefined \def \beditor#1{#1}\fi
\ifx \bpublisher  \undefined \def \bpublisher#1{#1}\fi
\ifx \bbtitle  \undefined \def \bbtitle#1{#1}\fi
\ifx \bedition  \undefined \def \bedition#1{#1}\fi
\ifx \bseriesno  \undefined \def \bseriesno#1{#1}\fi
\ifx \blocation  \undefined \def \blocation#1{#1}\fi
\ifx \bsertitle  \undefined \def \bsertitle#1{#1}\fi
\ifx \bsnm \undefined \def \bsnm#1{#1}\fi
\ifx \bsuffix \undefined \def \bsuffix#1{#1}\fi
\ifx \bparticle \undefined \def \bparticle#1{#1}\fi
\ifx \barticle \undefined \def \barticle#1{#1}\fi
\bibcommenthead
\ifx \bconfdate \undefined \def \bconfdate #1{#1}\fi
\ifx \botherref \undefined \def \botherref #1{#1}\fi
\ifx \url \undefined \def \url#1{\textsf{#1}}\fi
\ifx \bchapter \undefined \def \bchapter#1{#1}\fi
\ifx \bbook \undefined \def \bbook#1{#1}\fi
\ifx \bcomment \undefined \def \bcomment#1{#1}\fi
\ifx \oauthor \undefined \def \oauthor#1{#1}\fi
\ifx \citeauthoryear \undefined \def \citeauthoryear#1{#1}\fi
\ifx \endbibitem  \undefined \def \endbibitem {}\fi
\ifx \bconflocation  \undefined \def \bconflocation#1{#1}\fi
\ifx \arxivurl  \undefined \def \arxivurl#1{\textsf{#1}}\fi
\csname PreBibitemsHook\endcsname

\bibitem[\protect\citeauthoryear{{Rezzolla} et~al.}{2018}]{NSbook}
\begin{bbook}
\bauthor{\bsnm{{Rezzolla}}, \binits{L.}},
\bauthor{\bsnm{{Pizzochero}}, \binits{P.}},
\bauthor{\bsnm{{Jones}}, \binits{D.I.}},
\bauthor{\bsnm{{Rea}}, \binits{N.}}:
\bbtitle{{The Physics and Astrophysics of Neutron Stars}}.
\bsertitle{Astrophysics and Space Science Library},
vol. \bseriesno{457}
(\byear{2018}).
\doiurl{10.1007/978-3-319-97616-7}
\end{bbook}
\endbibitem

\bibitem[\protect\citeauthoryear{{Potekhin}}{1999}]{Potekhin99}
\begin{barticle}
\bauthor{\bsnm{{Potekhin}}, \binits{A.Y.}}:
\batitle{{Electron conduction in magnetized neutron star envelopes}}.
\bjtitle{\aap}
\bvolume{351},
\bfpage{787}--\blpage{797}
(\byear{1999})
\doiurl{10.48550/arXiv.astro-ph/9909100}
{\href{https://arxiv.org/abs/astro-ph/9909100}{{arXiv:astro-ph/9909100}}}
{[astro-ph]}
\end{barticle}
\endbibitem

\bibitem[\protect\citeauthoryear{{Bergeron} et~al.}{1992}]{Bergeron92}
\begin{barticle}
\bauthor{\bsnm{{Bergeron}}, \binits{P.}},
\bauthor{\bsnm{{Saffer}}, \binits{R.A.}},
\bauthor{\bsnm{{Liebert}}, \binits{J.}}:
\batitle{{A Spectroscopic Determination of the Mass Distribution of DA White
  Dwarfs}}.
\bjtitle{\apj}
\bvolume{394},
\bfpage{228}
(\byear{1992})
\doiurl{10.1086/171575}
\end{barticle}
\endbibitem

\bibitem[\protect\citeauthoryear{{Ventura}}{1973}]{Ventura73}
\begin{barticle}
\bauthor{\bsnm{{Ventura}}, \binits{J.}}:
\batitle{{Collision Frequency and Coulomb Scattering in an Intense Magnetic
  Field}}.
\bjtitle{\pra}
\bvolume{8}(\bissue{6}),
\bfpage{3021}--\blpage{3031}
(\byear{1973})
\doiurl{10.1103/PhysRevA.8.3021}
\end{barticle}
\endbibitem

\bibitem[\protect\citeauthoryear{{Oppenheimer}}{1928}]{Oppenheimer28}
\begin{barticle}
\bauthor{\bsnm{{Oppenheimer}}, \binits{J.R.}}:
\batitle{{On the Quantum Theory of Electronic Impacts}}.
\bjtitle{Physical Review}
\bvolume{32}(\bissue{3}),
\bfpage{361}--\blpage{376}
(\byear{1928})
\doiurl{10.1103/PhysRev.32.361}
\end{barticle}
\endbibitem

\bibitem[\protect\citeauthoryear{{Bray} and {Stelbovics}}{1992}]{Bray92}
\begin{barticle}
\bauthor{\bsnm{{Bray}}, \binits{I.}},
\bauthor{\bsnm{{Stelbovics}}, \binits{A.T.}}:
\batitle{{Convergent close-coupling calculations of electron-hydrogen
  scattering}}.
\bjtitle{\pra}
\bvolume{46}(\bissue{11}),
\bfpage{6995}--\blpage{7011}
(\byear{1992})
\doiurl{10.1103/PhysRevA.46.6995}
\end{barticle}
\endbibitem

\bibitem[\protect\citeauthoryear{Lattimer and Prakash}{2001}]{Lattimer01}
\begin{barticle}
\bauthor{\bsnm{Lattimer}, \binits{J.M.}},
\bauthor{\bsnm{Prakash}, \binits{M.}}:
\batitle{Neutron star structure and the equation of state}.
\bjtitle{The Astrophysical Journal}
\bvolume{550}(\bissue{1}),
\bfpage{426}
(\byear{2001})
\doiurl{10.1086/319702}
\end{barticle}
\endbibitem

\bibitem[\protect\citeauthoryear{{Lindblom}}{1992}]{Lindblom92}
\begin{barticle}
\bauthor{\bsnm{{Lindblom}}, \binits{L.}}:
\batitle{{Determining the Nuclear Equation of State from Neutron-Star Masses
  and Radii}}.
\bjtitle{\apj}
\bvolume{398},
\bfpage{569}
(\byear{1992})
\doiurl{10.1086/171882}
\end{barticle}
\endbibitem

\bibitem[\protect\citeauthoryear{{Heinke} et~al.}{2014}]{Heinke14}
\begin{barticle}
\bauthor{\bsnm{{Heinke}}, \binits{C.O.}},
\bauthor{\bsnm{{Cohn}}, \binits{H.N.}},
\bauthor{\bsnm{{Lugger}}, \binits{P.M.}},
\bauthor{\bsnm{{Webb}}, \binits{N.A.}},
\bauthor{\bsnm{{Ho}}, \binits{W.C.G.}},
\bauthor{\bsnm{{Anderson}}, \binits{J.}},
\bauthor{\bsnm{{Campana}}, \binits{S.}},
\bauthor{\bsnm{{Bogdanov}}, \binits{S.}},
\bauthor{\bsnm{{Haggard}}, \binits{D.}},
\bauthor{\bsnm{{Cool}}, \binits{A.M.}},
\bauthor{\bsnm{{Grindlay}}, \binits{J.E.}}:
\batitle{{Improved mass and radius constraints for quiescent neutron stars in
  {\ensuremath{\omega}} Cen and NGC 6397}}.
\bjtitle{\mnras}
\bvolume{444}(\bissue{1}),
\bfpage{443}--\blpage{456}
(\byear{2014})
\doiurl{10.1093/mnras/stu1449}
{\href{https://arxiv.org/abs/1406.1497}{{arXiv:1406.1497}}}
{[astro-ph.HE]}
\end{barticle}
\endbibitem

\bibitem[\protect\citeauthoryear{{{\"O}zel} and {Freire}}{2016}]{Ozel16}
\begin{barticle}
\bauthor{\bsnm{{{\"O}zel}}, \binits{F.}},
\bauthor{\bsnm{{Freire}}, \binits{P.}}:
\batitle{{Masses, Radii, and the Equation of State of Neutron Stars}}.
\bjtitle{\araa}
\bvolume{54},
\bfpage{401}--\blpage{440}
(\byear{2016})
\doiurl{10.1146/annurev-astro-081915-023322}
{\href{https://arxiv.org/abs/1603.02698}{{arXiv:1603.02698}}}
{[astro-ph.HE]}
\end{barticle}
\endbibitem

\bibitem[\protect\citeauthoryear{{Bogdanov} et~al.}{2019a}]{NICER1}
\begin{barticle}
\bauthor{\bsnm{{Bogdanov}}, \binits{S.}},
\bauthor{\bsnm{{Guillot}}, \binits{S.}},
\bauthor{\bsnm{{Ray}}, \binits{P.S.}},
\bauthor{\bsnm{{Wolff}}, \binits{M.T.}},
\bauthor{\bsnm{{Chakrabarty}}, \binits{D.}},
\bauthor{\bsnm{{Ho}}, \binits{W.C.G.}},
\bauthor{\bsnm{{Kerr}}, \binits{M.}},
\bauthor{\bsnm{{Lamb}}, \binits{F.K.}},
\bauthor{\bsnm{{Lommen}}, \binits{A.}},
\bauthor{\bsnm{{Ludlam}}, \binits{R.M.}},
\bauthor{\bsnm{{Milburn}}, \binits{R.}},
\bauthor{\bsnm{{Montano}}, \binits{S.}},
\bauthor{\bsnm{{Miller}}, \binits{M.C.}},
\bauthor{\bsnm{{Baub{\"o}ck}}, \binits{M.}},
\bauthor{\bsnm{{{\"O}zel}}, \binits{F.}},
\bauthor{\bsnm{{Psaltis}}, \binits{D.}},
\bauthor{\bsnm{{Remillard}}, \binits{R.A.}},
\bauthor{\bsnm{{Riley}}, \binits{T.E.}},
\bauthor{\bsnm{{Steiner}}, \binits{J.F.}},
\bauthor{\bsnm{{Strohmayer}}, \binits{T.E.}},
\bauthor{\bsnm{{Watts}}, \binits{A.L.}},
\bauthor{\bsnm{{Wood}}, \binits{K.S.}},
\bauthor{\bsnm{{Zeldes}}, \binits{J.}},
\bauthor{\bsnm{{Enoto}}, \binits{T.}},
\bauthor{\bsnm{{Okajima}}, \binits{T.}},
\bauthor{\bsnm{{Kellogg}}, \binits{J.W.}},
\bauthor{\bsnm{{Baker}}, \binits{C.}},
\bauthor{\bsnm{{Markwardt}}, \binits{C.B.}},
\bauthor{\bsnm{{Arzoumanian}}, \binits{Z.}},
\bauthor{\bsnm{{Gendreau}}, \binits{K.C.}}:
\batitle{{Constraining the Neutron Star Mass-Radius Relation and Dense Matter
  Equation of State with NICER. I. The Millisecond Pulsar X-Ray Data Set}}.
\bjtitle{\apjl}
\bvolume{887}(\bissue{1}),
\bfpage{25}
(\byear{2019})
\doiurl{10.3847/2041-8213/ab53eb}
{\href{https://arxiv.org/abs/1912.05706}{{arXiv:1912.05706}}}
{[astro-ph.HE]}
\end{barticle}
\endbibitem

\bibitem[\protect\citeauthoryear{{Bogdanov} et~al.}{2019b}]{NICER2}
\begin{barticle}
\bauthor{\bsnm{{Bogdanov}}, \binits{S.}},
\bauthor{\bsnm{{Lamb}}, \binits{F.K.}},
\bauthor{\bsnm{{Mahmoodifar}}, \binits{S.}},
\bauthor{\bsnm{{Miller}}, \binits{M.C.}},
\bauthor{\bsnm{{Morsink}}, \binits{S.M.}},
\bauthor{\bsnm{{Riley}}, \binits{T.E.}},
\bauthor{\bsnm{{Strohmayer}}, \binits{T.E.}},
\bauthor{\bsnm{{Tung}}, \binits{A.K.}},
\bauthor{\bsnm{{Watts}}, \binits{A.L.}},
\bauthor{\bsnm{{Dittmann}}, \binits{A.J.}},
\bauthor{\bsnm{{Chakrabarty}}, \binits{D.}},
\bauthor{\bsnm{{Guillot}}, \binits{S.}},
\bauthor{\bsnm{{Arzoumanian}}, \binits{Z.}},
\bauthor{\bsnm{{Gendreau}}, \binits{K.C.}}:
\batitle{{Constraining the Neutron Star Mass-Radius Relation and Dense Matter
  Equation of State with NICER. II. Emission from Hot Spots on a Rapidly
  Rotating Neutron Star}}.
\bjtitle{\apjl}
\bvolume{887}(\bissue{1}),
\bfpage{26}
(\byear{2019})
\doiurl{10.3847/2041-8213/ab5968}
{\href{https://arxiv.org/abs/1912.05707}{{arXiv:1912.05707}}}
{[astro-ph.HE]}
\end{barticle}
\endbibitem

\bibitem[\protect\citeauthoryear{{Ruder} et~al.}{1994}]{Ruder94}
\begin{bbook}
\bauthor{\bsnm{{Ruder}}, \binits{H.}},
\bauthor{\bsnm{{Wunner}}, \binits{G.}},
\bauthor{\bsnm{{Herold}}, \binits{H.}},
\bauthor{\bsnm{{Geyer}}, \binits{F.}}:
\bbtitle{{Atoms in Strong Magnetic Fields. Quantum Mechanical Treatment and
  Applications in Astrophysics and Quantum Chaos}},
(\byear{1994})
\end{bbook}
\endbibitem

\bibitem[\protect\citeauthoryear{{Ziman}}{1954}]{Ziman54}
\begin{barticle}
\bauthor{\bsnm{{Ziman}}, \binits{J.M.}}:
\batitle{{The Electrical and Thermal Conductivities of Monovalent Metals}}.
\bjtitle{Proceedings of the Royal Society of London Series A}
\bvolume{226}(\bissue{1167}),
\bfpage{436}--\blpage{454}
(\byear{1954})
\doiurl{10.1098/rspa.1954.0267}
\end{barticle}
\endbibitem

\bibitem[\protect\citeauthoryear{{Baalrud} and {Daligault}}{2017}]{Baalrud17}
\begin{barticle}
\bauthor{\bsnm{{Baalrud}}, \binits{S.D.}},
\bauthor{\bsnm{{Daligault}}, \binits{J.}}:
\batitle{{Transport regimes spanning magnetization-coupling phase space}}.
\bjtitle{\pre}
\bvolume{96}(\bissue{4}),
\bfpage{043202}
(\byear{2017})
\doiurl{10.1103/PhysRevE.96.043202}
{\href{https://arxiv.org/abs/1709.05420}{{arXiv:1709.05420}}}
{[physics.plasm-ph]}
\end{barticle}
\endbibitem

\bibitem[\protect\citeauthoryear{{Braginskii}}{1965}]{Braginskii}
\begin{barticle}
\bauthor{\bsnm{{Braginskii}}, \binits{S.I.}}:
\batitle{{Transport Processes in a Plasma}}.
\bjtitle{Reviews of Plasma Physics}
\bvolume{1},
\bfpage{205}
(\byear{1965})
\end{barticle}
\endbibitem

\bibitem[\protect\citeauthoryear{{Lee} and {More}}{1984}]{LeeMore84}
\begin{barticle}
\bauthor{\bsnm{{Lee}}, \binits{Y.T.}},
\bauthor{\bsnm{{More}}, \binits{R.M.}}:
\batitle{{An electron conductivity model for dense plasmas}}.
\bjtitle{Physics of Fluids}
\bvolume{27}(\bissue{5}),
\bfpage{1273}--\blpage{1286}
(\byear{1984})
\doiurl{10.1063/1.864744}
\end{barticle}
\endbibitem

\bibitem[\protect\citeauthoryear{{Baranger}}{1958}]{Baranger58c}
\begin{barticle}
\bauthor{\bsnm{{Baranger}}, \binits{M.}}:
\batitle{{General Impact Theory of Pressure Broadening}}.
\bjtitle{Physical Review}
\bvolume{112}(\bissue{3}),
\bfpage{855}--\blpage{865}
(\byear{1958})
\doiurl{10.1103/PhysRev.112.855}
\end{barticle}
\endbibitem

\bibitem[\protect\citeauthoryear{{Fano}}{1963}]{Fano63}
\begin{barticle}
\bauthor{\bsnm{{Fano}}, \binits{U.}}:
\batitle{{Pressure Broadening as a Prototype of Relaxation}}.
\bjtitle{Physical Review}
\bvolume{131}(\bissue{1}),
\bfpage{259}--\blpage{268}
(\byear{1963})
\doiurl{10.1103/PhysRev.131.259}
\end{barticle}
\endbibitem

\bibitem[\protect\citeauthoryear{{Gomez} et~al.}{2022}]{Gomez22}
\begin{barticle}
\bauthor{\bsnm{{Gomez}}, \binits{T.A.}},
\bauthor{\bsnm{{Nagayama}}, \binits{T.}},
\bauthor{\bsnm{{Cho}}, \binits{P.B.}},
\bauthor{\bsnm{{Kilcrease}}, \binits{D.P.}},
\bauthor{\bsnm{{Fontes}}, \binits{C.J.}},
\bauthor{\bsnm{{Zammit}}, \binits{M.C.}}:
\batitle{{Introduction to spectral line shape theory}}.
\bjtitle{Journal of Physics B Atomic Molecular Physics}
\bvolume{55}(\bissue{3}),
\bfpage{034002}
(\byear{2022})
\doiurl{10.1088/1361-6455/ac4f31}
\end{barticle}
\endbibitem

\bibitem[\protect\citeauthoryear{{Gomez} et~al.}{2023}]{Gomez23}
\begin{barticle}
\bauthor{\bsnm{{Gomez}}, \binits{T.A.}},
\bauthor{\bsnm{{Zammit}}, \binits{M.C.}},
\bauthor{\bsnm{{Fontes}}, \binits{C.J.}},
\bauthor{\bsnm{{White}}, \binits{J.R.}}:
\batitle{{A Quantum-mechanical Treatment of Electron Broadening in Strong
  Magnetic Fields}}.
\bjtitle{\apj}
\bvolume{951}(\bissue{2}),
\bfpage{143}
(\byear{2023})
\doiurl{10.3847/1538-4357/acda28}
\end{barticle}
\endbibitem

\bibitem[\protect\citeauthoryear{{XRISM Science Team}}{2022}]{XRISM}
\begin{botherref}
\oauthor{\bsnm{{XRISM Science Team}}}:
{XRISM Quick Reference}.
arXiv e-prints,
2202--05399
(2022)
\doiurl{10.48550/arXiv.2202.05399}
{\href{https://arxiv.org/abs/2202.05399}{{arXiv:2202.05399}}}
{[astro-ph.IM]}
\end{botherref}
\endbibitem

\bibitem[\protect\citeauthoryear{Smith et~al.}{2017}]{Smith17_arcus}
\begin{bchapter}
\bauthor{\bsnm{Smith}, \binits{R.K.}},
\bauthor{\bsnm{Abraham}, \binits{M.}},
\bauthor{\bsnm{Allured}, \binits{R.}},
\bauthor{\bsnm{Bautz}, \binits{M.}},
\bauthor{\bsnm{Bookbinder}, \binits{J.}},
\bauthor{\bsnm{Bregman}, \binits{J.}},
\bauthor{\bsnm{Brenneman}, \binits{L.}},
\bauthor{\bsnm{Brickhouse}, \binits{N.S.}},
\bauthor{\bsnm{Burrows}, \binits{D.}},
\bauthor{\bsnm{Burwitz}, \binits{V.}}, \betal:
\bctitle{Arcus: exploring the formation and evolution of clusters, galaxies,
  and stars}.
In: \bbtitle{UV, X-Ray, and Gamma-Ray Space Instrumentation for Astronomy XX},
vol. \bseriesno{10397},
pp. \bfpage{194}--\blpage{204}
(\byear{2017}).
\bcomment{SPIE}
\end{bchapter}
\endbibitem

\bibitem[\protect\citeauthoryear{Ferri et~al.}{2022}]{Ferri21}
\begin{barticle}
\bauthor{\bsnm{Ferri}, \binits{S.}},
\bauthor{\bsnm{Peyrusse}, \binits{O.}},
\bauthor{\bsnm{Calisti}, \binits{A.}}:
\batitle{Stark-zeeman line-shape model for multi-electron radiators in hot
  dense plasmas subjected to large magnetic fields}.
\bjtitle{Matter and Radiation at Extremes}
\bvolume{7},
\bfpage{015901}
(\byear{2022})
\doiurl{10.1063/5.0058552}
\end{barticle}
\endbibitem

\bibitem[\protect\citeauthoryear{{Lafleur} and {Baalrud}}{2019}]{Lafleur19}
\begin{barticle}
\bauthor{\bsnm{{Lafleur}}, \binits{T.}},
\bauthor{\bsnm{{Baalrud}}, \binits{S.D.}}:
\batitle{{Transverse force induced by a magnetized wake}}.
\bjtitle{Plasma Physics and Controlled Fusion}
\bvolume{61}(\bissue{12}),
\bfpage{125004}
(\byear{2019})
\doiurl{10.1088/1361-6587/ab45d4}
{\href{https://arxiv.org/abs/1909.03121}{{arXiv:1909.03121}}}
{[physics.plasm-ph]}
\end{barticle}
\endbibitem

\bibitem[\protect\citeauthoryear{{Ho} et~al.}{2003}]{Ho03}
\begin{barticle}
\bauthor{\bsnm{{Ho}}, \binits{W.C.G.}},
\bauthor{\bsnm{{Lai}}, \binits{D.}},
\bauthor{\bsnm{{Potekhin}}, \binits{A.Y.}},
\bauthor{\bsnm{{Chabrier}}, \binits{G.}}:
\batitle{{Atmospheres and Spectra of Strongly Magnetized Neutron Stars. III.
  Partially Ionized Hydrogen Models}}.
\bjtitle{\apj}
\bvolume{599}(\bissue{2}),
\bfpage{1293}--\blpage{1301}
(\byear{2003})
\doiurl{10.1086/379507}
{\href{https://arxiv.org/abs/astro-ph/0309261}{{arXiv:astro-ph/0309261}}}
{[astro-ph]}
\end{barticle}
\endbibitem

\bibitem[\protect\citeauthoryear{{G{\"u}ver} et~al.}{2011}]{Guver11}
\begin{barticle}
\bauthor{\bsnm{{G{\"u}ver}}, \binits{T.}},
\bauthor{\bsnm{{G{\"o}{\v{g}}{\"u}{\c{s}}}}, \binits{E.}},
\bauthor{\bsnm{{{\"O}zel}}, \binits{F.}}:
\batitle{{A magnetar strength surface magnetic field for the slowly spinning
  down SGR 0418+5729}}.
\bjtitle{\mnras}
\bvolume{418}(\bissue{4}),
\bfpage{2773}--\blpage{2778}
(\byear{2011})
\doiurl{10.1111/j.1365-2966.2011.19677.x}
{\href{https://arxiv.org/abs/1103.3024}{{arXiv:1103.3024}}}
{[astro-ph.HE]}
\end{barticle}
\endbibitem

\bibitem[\protect\citeauthoryear{{Mori} and {Hailey}}{2006}]{Mori06}
\begin{barticle}
\bauthor{\bsnm{{Mori}}, \binits{K.}},
\bauthor{\bsnm{{Hailey}}, \binits{C.J.}}:
\batitle{{Detailed Atmosphere Modeling for the Neutron Star 1E1207.4-5209:
  Evidence of Oxygen/Neon Atmosphere}}.
\bjtitle{\apj}
\bvolume{648}(\bissue{2}),
\bfpage{1139}--\blpage{1155}
(\byear{2006})
\doiurl{10.1086/506008}
{\href{https://arxiv.org/abs/astro-ph/0301161}{{arXiv:astro-ph/0301161}}}
{[astro-ph]}
\end{barticle}
\endbibitem

\bibitem[\protect\citeauthoryear{{Ho} and {Heinke}}{2009}]{Ho09}
\begin{barticle}
\bauthor{\bsnm{{Ho}}, \binits{W.C.G.}},
\bauthor{\bsnm{{Heinke}}, \binits{C.O.}}:
\batitle{{A neutron star with a carbon atmosphere in the Cassiopeia A supernova
  remnant}}.
\bjtitle{\nat}
\bvolume{462}(\bissue{7269}),
\bfpage{71}--\blpage{73}
(\byear{2009})
\doiurl{10.1038/nature08525}
{\href{https://arxiv.org/abs/0911.0672}{{arXiv:0911.0672}}}
{[astro-ph.HE]}
\end{barticle}
\endbibitem

\bibitem[\protect\citeauthoryear{{Alford} and {Halpern}}{2023}]{Alford23}
\begin{barticle}
\bauthor{\bsnm{{Alford}}, \binits{J.A.J.}},
\bauthor{\bsnm{{Halpern}}, \binits{J.P.}}:
\batitle{{Do Central Compact Objects have Carbon Atmospheres?}}
\bjtitle{\apj}
\bvolume{944}(\bissue{1}),
\bfpage{36}
(\byear{2023})
\doiurl{10.3847/1538-4357/acaf55}
{\href{https://arxiv.org/abs/2302.05893}{{arXiv:2302.05893}}}
{[astro-ph.HE]}
\end{barticle}
\endbibitem

\bibitem[\protect\citeauthoryear{{Clark}}{1983}]{Clark83}
\begin{barticle}
\bauthor{\bsnm{{Clark}}, \binits{C.W.}}:
\batitle{{Low-energy electron-atom scattering in a magnetic field}}.
\bjtitle{\pra}
\bvolume{28}(\bissue{1}),
\bfpage{83}--\blpage{90}
(\byear{1983})
\doiurl{10.1103/PhysRevA.28.83}
\end{barticle}
\endbibitem

\bibitem[\protect\citeauthoryear{{Lippmann} and
  {Schwinger}}{1950}]{LippmannSchwinger50}
\begin{barticle}
\bauthor{\bsnm{{Lippmann}}, \binits{B.A.}},
\bauthor{\bsnm{{Schwinger}}, \binits{J.}}:
\batitle{{Variational Principles for Scattering Processes. I}}.
\bjtitle{Physical Review}
\bvolume{79}(\bissue{3}),
\bfpage{469}--\blpage{480}
(\byear{1950})
\doiurl{10.1103/PhysRev.79.469}
\end{barticle}
\endbibitem

\bibitem[\protect\citeauthoryear{{Bethe} and {Salpeter}}{1957}]{BetheSalpeter}
\begin{bbook}
\bauthor{\bsnm{{Bethe}}, \binits{H.A.}},
\bauthor{\bsnm{{Salpeter}}, \binits{E.E.}}:
\bbtitle{{Quantum Mechanics of One- and Two-Electron Atoms}},
(\byear{1957})
\end{bbook}
\endbibitem

\bibitem[\protect\citeauthoryear{{Mott} and {Massey}}{1949}]{MottMassey}
\begin{bbook}
\bauthor{\bsnm{{Mott}}, \binits{N.F.}},
\bauthor{\bsnm{{Massey}}, \binits{H.S.W.}}:
\bbtitle{{The Theory of Atomic Collisions}},
(\byear{1949})
\end{bbook}
\endbibitem

\bibitem[\protect\citeauthoryear{{Moiseiwitsch} and
  {Smith}}{1968}]{Moiseiwitsch68}
\begin{barticle}
\bauthor{\bsnm{{Moiseiwitsch}}, \binits{B.L.}},
\bauthor{\bsnm{{Smith}}, \binits{S.J.}}:
\batitle{{Electron Impact Excitation of Atoms}}.
\bjtitle{Reviews of Modern Physics}
\bvolume{40}(\bissue{2}),
\bfpage{238}--\blpage{353}
(\byear{1968})
\doiurl{10.1103/RevModPhys.40.238}
\end{barticle}
\endbibitem

\bibitem[\protect\citeauthoryear{{Gomez} et~al.}{2021}]{Gomez21}
\begin{barticle}
\bauthor{\bsnm{{Gomez}}, \binits{T.A.}},
\bauthor{\bsnm{{Nagayama}}, \binits{T.}},
\bauthor{\bsnm{{Cho}}, \binits{P.B.}},
\bauthor{\bsnm{{Zammit}}, \binits{M.C.}},
\bauthor{\bsnm{{Fontes}}, \binits{C.J.}},
\bauthor{\bsnm{{Kilcrease}}, \binits{D.P.}},
\bauthor{\bsnm{{Bray}}, \binits{I.}},
\bauthor{\bsnm{{Hubeny}}, \binits{I.}},
\bauthor{\bsnm{{Dunlap}}, \binits{B.H.}},
\bauthor{\bsnm{{Montgomery}}, \binits{M.H.}},
\bauthor{\bsnm{{Winget}}, \binits{D.E.}}:
\batitle{{All-Order Full-Coulomb Quantum Spectral Line-Shape Calculations}}.
\bjtitle{\prl}
\bvolume{127}(\bissue{23}),
\bfpage{235001}
(\byear{2021})
\doiurl{10.1103/PhysRevLett.127.235001}
\end{barticle}
\endbibitem

\bibitem[\protect\citeauthoryear{{Bray}}{1994}]{Bray94}
\begin{barticle}
\bauthor{\bsnm{{Bray}}, \binits{I.}}:
\batitle{{Convergent close-coupling method for the calculation of electron
  scattering on hydrogenlike targets}}.
\bjtitle{\pra}
\bvolume{49}(\bissue{2}),
\bfpage{1066}--\blpage{1082}
(\byear{1994})
\doiurl{10.1103/PhysRevA.49.1066}
\end{barticle}
\endbibitem

\bibitem[\protect\citeauthoryear{{Gomez} et~al.}{2020}]{Gomez20}
\begin{barticle}
\bauthor{\bsnm{{Gomez}}, \binits{T.A.}},
\bauthor{\bsnm{{Nagayama}}, \binits{T.}},
\bauthor{\bsnm{{Fontes}}, \binits{C.J.}},
\bauthor{\bsnm{{Kilcrease}}, \binits{D.P.}},
\bauthor{\bsnm{{Hansen}}, \binits{S.B.}},
\bauthor{\bsnm{{Zammit}}, \binits{M.C.}},
\bauthor{\bsnm{{Fursa}}, \binits{D.V.}},
\bauthor{\bsnm{{Kadyrov}}, \binits{A.S.}},
\bauthor{\bsnm{{Bray}}, \binits{I.}}:
\batitle{{Effect of Electron Capture on Spectral Line Broadening in Hot Dense
  Plasmas}}.
\bjtitle{\prl}
\bvolume{124}(\bissue{5}),
\bfpage{055003}
(\byear{2020})
\doiurl{10.1103/PhysRevLett.124.055003}
\end{barticle}
\endbibitem

\bibitem[\protect\citeauthoryear{{Johnson} et~al.}{1983}]{Johnson83}
\begin{barticle}
\bauthor{\bsnm{{Johnson}}, \binits{B.R.}},
\bauthor{\bsnm{{Hirschfelder}}, \binits{J.O.}},
\bauthor{\bsnm{{Yang}}, \binits{K.-H.}}:
\batitle{{Interaction of atoms, molecules, and ions with constant electric and
  magnetic fields}}.
\bjtitle{Reviews of Modern Physics}
\bvolume{55}(\bissue{1}),
\bfpage{109}--\blpage{153}
(\byear{1983})
\doiurl{10.1103/RevModPhys.55.109}
\end{barticle}
\endbibitem

\bibitem[\protect\citeauthoryear{{Pavlov} and {Meszaros}}{1993}]{PM93}
\begin{barticle}
\bauthor{\bsnm{{Pavlov}}, \binits{G.G.}},
\bauthor{\bsnm{{Meszaros}}, \binits{P.}}:
\batitle{{Finite-Velocity Effects on Atoms in Strong Magnetic Fields and
  Implications for Neutron Star Atmospheres}}.
\bjtitle{\apj}
\bvolume{416},
\bfpage{752}
(\byear{1993})
\doiurl{10.1086/173274}
\end{barticle}
\endbibitem

\bibitem[\protect\citeauthoryear{{McQuarrie}}{2003}]{McQuarrie}
\begin{bbook}
\bauthor{\bsnm{{McQuarrie}}, \binits{D.A.}}:
\bbtitle{{Statistical Mechanics}}.
\bpublisher{{University Science Books}}, \blocation{???}
(\byear{2003})
\end{bbook}
\endbibitem

\end{thebibliography}
\end{document}